# Synthesized inverse-perovskites Sc$_3$In$X$ ($X$ = B, C, N): A theoretical investigation

M.A. Hossain[1], M.S. Ali[2], F. Parvin[2], A.K.M.A. Islam[2*]

[1]Department of Physics, Mawlana Bhashani Science and Technology University, Tangail-1902, Bangladesh

[2]Department of Physics, Rajshahi University, Rajshahi, Bangladesh

**Abstract**

We present first-principles density functional theory (DFT) investigations of mechanical, thermodynamic and optical properties of synthesized inverse-perovskites Sc$_3$In$X$ ($X$ = B, C, N). The elastic constants at zero pressure and temperature are calculated and the anisotropic behavior of the compounds is illustrated. All the three materials are shown to be brittle in nature. The computed Peierls stress, approximately 3 to 5 times larger than of a selection of MAX phases, show that dislocation movement may follow but with much reduced occurrences compared to these MAX phases. The Mulliken bonding population and charge density maps show stronger covalency between Sc and $X$ atoms compared with Sc-Sc bond. The Vickers hardness values of Sc$_3$In$X$ are predicted to be between 3.03 and 3.88 GPa. The Fermi surfaces of Sc$_3$In$X$ contain both hole- and electron-like topology which changes as one replaces B with C or N. The bulk modulus, specific heats, thermal expansion coefficient, and Debye temperature are calculated as a function both temperature and pressure using the quasi-harmonic Debye model with phononic effects. The results so obtained are analysed in comparison to the characteristics of other related compounds. Moreover optical functions are calculated and discussed for the first time. The reflectivity is found to be high in the IR-UV regions up to ~ 10.7 eV (Sc$_3$InB, Sc$_3$InC) and 12.3 eV (Sc$_3$InN), thus showing promise as good coating materials.

## 1. Introduction

The ternary borides, carbides or nitrides with cubic inverse-perovskite structure belong to a series of materials with general formula M$_3$A$X$, where *M* is a transition metal, *A* is a main group element, where $X$ is boron, carbon or nitrogen [1]. In this structure, *A* atom is located at the origin, $X$ is in the corner position of the cube and *M* atoms are located at the face-centered positions. It has been found that these anti-perovskite compounds display interesting properties such as nearly zero temperature coefficient of resistivity [2], giant magnetoresistance [3], and depending on their chemical composition, can also exhibit a wide variety of physical properties ranging from semiconducting to magnetic and superconducting properties [4-9]. The inverse-perovskite compounds could also be more important regarding to other properties such as hardness and diversity of electronic properties [10 -12].

The discovery of superconductivity in the inverse-perovskite Ni$_3$MgC compound [13] has generated interest in the inverse-perovskite family. Takei *et al.* [14] reported that among other intermetallic perovskite similar to Ni$_3$MgC, only one compound, YRh$_3$B is a superconductor with $T_c$ = 0.76 K. Wiendlocha *et al.* [15] have investigated theoretically the electronic properties and predicted that scandium based borides, Sc$_3$XB (X = Tl, In, Ga, Al) are superconductors with transition temperature 15, 12, 10 and 7.5 K, respectively. They have also studied the effect of vacancy on the boron site in Sc$_3$XB and this effect is very important for superconductivity. Among the four scandium-based borides the synthesis of Tl- and In-containing cubic perovskite have been reported by Holleck [16] while Ga- and Al- compounds are hypothetical.

---

[*] Corresponding author. Tel.: +88 0721 750980; fax: +88 0721 750064.
  *E-mail address*: azi46@ru.ac.bd (A.K.M.A. Islam).

Sc-based cubic inverse-perovskite carbides (CIPCs) have been synthesized by Gesing and collaborators [17]. Shein *et al.* [18] have studied the electronic properties of Sc$_3$AlC using a first-principles method and after few years Medkour *et al.* [19] reported a first-principles study on the structural, elastic and electronic properties of Sc$_3$AlC at zero pressure. Recently Haddadi *et al.* [20] have investigated structural, electronic, elastic and thermodynamic properties of the ternary scandium based inverse perovskite carbides Sc$_3$AC (A = Al, Ga, In and Tl). At about the same time Kanchan *et al.* [21] have also investigated the electronic structure and mechanical properties of these compoundsThe preparation, crystal structure and some physical properties of the inverse perovskite Sc$_3$InN have been reported by Kirchner *et al*. [22].

Sc$_3$AlN with the inverse-perovskite structure was synthesized for the first time by Höglund *et al.* [23]. Mattesini *et al.* [24] theoretically predicted Sc$_3$GaN with the inverse-perovskite structure by means of *ab initio* calculations. Electronic properties and structural stability of the ternary inverse-perovskites Sc$_3$BN, Sc$_3$AlN, Sc$_3$GaN and Sc$_3$InN have been thoroughly investigated using *ab initio* calculations [23-26]. Recently mechanical and thermodynamic stability of the isoelectronic ternary inverse perovskite Sc$_3$EN (E = B, Al, Ga, In) has been studied [25]. Among the various Sc-based inverse perovskite nitrides, Sc$_3$InN and Sc$_3$AlN phases have been synthesized [22, 23] and Sc$_3$GaN and Sc$_3$TlN phases are hypothetical. Mattesini *et al.* [24] have studied structural, electronic and elastic properties of Sc-based Al, Ga and In nitrides by first-principles method.

It is thus obvious from the above discussion that besides synthesis no other experimental work has been performed on Sc$_3$XB (X =Tl, In) [16], Sc$_3$AC (A = Al, Ga, In, Tl) [17, 27, 28] and Sc$_3$EN (E = Al, In,) [22, 23] compounds. For Sc$_3$AlN compound the electronic structure and chemical bonding in comparison to ScN and Sc metal have been investigated by bulk-sensitive soft x-ray emission spectroscopy [26] and the measured emission spectra are compared with the calculated spectra using first principles density functional theory including dipole transition matrix elements.

Sc$_3$InX (X = B, C, N) compounds with In atoms are important member of the ternary borides, carbides or nitrides with cubic inverse-perovskite structure. From our previous discussions we see that there is a dearth of information on these cubic inverse perovskites. To the best of our knowledge, mechanical, thermal and optical properties of Sc$_3$InB compound have not yet been discussed in literature. The optical properties, Peierls stress that measure the dislocation in a crystal, and Vikers hardness of Sc$_3$InC are also not available. Furthermore there is no report in literature on optical properties, Peierls stress, Vikers hardness and thermal properties of Sc$_3$InN. The specific heat of a material is one of the most important thermal properties indicating its heat retention or loss ability. The microscopic thermodynamic properties are closely related to the microscopic dynamics of atoms. In fact the thermodynamic properties of solids include a variety of other properties and phenomena such as bulk modulus, thermal expansion coefficient, and Debye temperature. On the other hand the optical properties of solids provide an important tool for studying energy band structure, impurity levels, excitons, localized defects, lattice vibrations, and certain magnetic excitations. The optical conductivity or the dielectric function indicates a response of a system of electrons to an applied field. Thus there is a need to deal with all these issues which will be covered in the present paper and a discussion and analysis will be made in comparison with the results of other relevant compounds.

## 2. Computational details

The plane-wave pseudopotential based on density functional theory (DFT) as implemented in CASTEP code [29] has been used to calculate the zero-temperature energy. The electronic exchange-correlation energy is treated under the generalized gradient approximation (GGA) in the scheme of Perdew-Burke-Ernzerhof (PBE) [30]. The ultrasoft Vanderbilt-type pseudopotentials for Sc, In, B, C, and N atoms have been employed to represent the interactions between ion and electron [31]. The calculations use a planewave cutoff energy 500 eV for all cases. For the sampling of the Brillouin zone, 8×8×8 $k$-point grids generated according to the Monkhorst-Pack scheme [32] are utilized. These parameters are found to be sufficient to lead to convergence of total energy giving geometrical configuration. Geometry optimization is achieved using convergence thresholds of $5\times10^{-6}$ eV/atom for the total energy, 0.01 eV/Å for the maximum force, 0.02 GPa for the maximum stress and $5\times10^{-4}$ Å for maximum displacement. The reciprocal space integrations are performed by using the tetrahedron method with a $k$-mesh of 20 $k$-points in the irreducible wedge of Brillouin zone (BZ). The total energy is converged to within 0.1mRy/unit cell during the self consistency cycle.

We utilized the quasi-harmonic Debye model to calculate the thermodynamic properties including the bulk modulus, thermal expansion coefficient, specific heats, and Debye temperature at any temperature and pressure. The DFT calculated energy-volume data at $T = 0$ K, $P = 0$ GPa and the Birch-Murnaghan EOS [34] are used for the purpose. The detailed procedures can be found in Ref. [33, 35].

## 3. Results and discussion

### 3.1. Mechanical properties

The scandium based ternary boride, carbide and nitide are cubic inverse perovskites with space group $Pm\overline{3}m$, the unit cell of which is displayed in Fig. 1. The equilibrium crystal structure of $Sc_3InX$ ($X$ = B, C, and N) is first obtained by minimizing the total energy. The spin-polarized calculations of the compounds converged to the nonmagnetic ground state. The relevant optimized lattice parameters are shown in Table 1 along with experimental data and other theoretical results [15-17, 20-22, 24, 28]. It is seen that the calculated lattice parameters for $Sc_3InB$ deviate by 2.8% from the experimental value obtained by Holleck [16]. For $Sc_3InC$ and $Sc_3InN$, the deviations of the calculated lattice parameters from the experimental values are 0.2% and 0.4%, respectively.

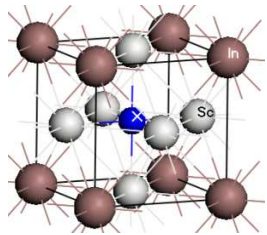

**Fig. 1.** Equilibrium crystal structure of $Sc_3InX$ ($X$ = B, C and N).

The elastic constants are essential parameters for illustrating the mechanical properties of materials that determine the responses of the compounds to applied strain or stress. Hence efforts have been taken to calculate elastic constants of cubic inverse perovskite Sc$_3$In$X$ phases. There are three independent elastic constants ($C_{11}$, $C_{12}$ and $C_{44}$) for cubic lattices. The elastic constant tensor of cubic inverse perovskite Sc$_3$In$X$ phases is reported in Table 1 along with available computed elastic constants. Using the second order elastic constants, the bulk modulus $B$, shear modulus $G$, Young's modulus $E$, and Poisson's ratio $v$ at zero pressure are calculated and illustrated in Table 1. The elastic constants of Sc$_3$InB are calculated for the first time. The calculated elastic constants for Sc$_3$InC and Sc$_3$InN are compared with other theoretical results and it is found that our results agree well with these. It is also observed that $C_{11}$ decreases and $C_{12}$, $C_{44}$ increases when we replace boron by carbon or nitrogen in Sc$_3$In$X$ phases.

**Table 1.** Calculated lattice parameters $a$ (Å), elastic constants $C_{ij}$ (GPa), bulk moduli $B$ (GPa), shear moduli $G$ (GPa), Young's moduli $E$ (GPa), Poisson's ratio $v$, elastic anisotropy $A$, Debye temeperature $\Theta_D$ (K), Burgers vector $b$ (Å) and interlayer distance $d$ (Å), and Peierls stress $\sigma_p$ (GPa) for Sc$_3$In$X$ phases ($X$=B, C, N).

| Structure | $a$ | $C_{11}$ | $C_{12}$ | $C_{44}$ | $B$ | $G$ | $E$ | $v$ | $A$ | $\Theta_B$ | $b$ | $d$ | $\sigma_P$ |
|---|---|---|---|---|---|---|---|---|---|---|---|---|---|
| Sc$_3$InB | 4.687 (calc.) | 163.5 | 36.2 | 50.9 | 78.6 | 55.7 | 135.1 | 0.214 | 0.80 | 437 | 4.687 | 2.344 | 3.06 |
| | 4.556 (theo)[a] | | | | | | | | | | | | |
| | 4.561 (expt)[b] | | | | | | | | | | | | |
| Sc$_3$InC | 4.553 (calc.) | 217.8 | 42.7 | 82.0 | 101.1 | 84.2 | 197.7 | 0.174 | 0.94 | 526 | 4.553 | 2.277 | 4.2 |
| | 4.5368 (expt)[c] | - | - | - | - | - | - | - | - | - | - | - | - |
| | 4.553 (theo)[d] | 211.3 | 41.1 | 80.6 | 97.3 | 82.4 | 193.0 | 0.171 | 0.95 | 521 | - | - | - |
| | 4.56 (theo)[e] | 213 | 38 | 81 | 96.1 | 82.7 | 193 | 0.170 | 0.93 | 522 | - | - | - |
| | 4.54 (expt)[h] | - | - | - | - | - | - | - | - | - | - | - | - |
| Sc$_3$InN | 4.47 (calc.) | 208.7 | 52.8 | 83.7 | 104.7 | 81.4 | 193.9 | 0.192 | 1.07 | 512 | 4.47 | 2.235 | 5.31 |
| | 4.47 (theo)[e] | 210 | 54 | 80 | 106.1 | 80.4 | 193 | 0.200 | 1.06 | 509 | - | - | - |
| | 4.45 (expt)[f] | - | - | - | - | - | - | - | - | - | - | - | - |
| | 4.411 (theo)[g] | 238.6 | 54.3 | 90.8 | 115.7 | 91.3 | 216.9 | 0.188 | 0.98 | 538 | - | - | - |

a: [15]
b: [16]
c: [17]
d: [20]
e: [21]
f: [22]
g: [24]
h. [28]

The calculated single crystal elastic constants of all the compounds satisfy the criterion of stability [21]. The idea of elastic anisotropy is essential in various applications such as dislocation dynamics, development of microcracks, phase transformations and other geophysical applications. Further anisotropy is reported to influence the nanoscale precursor textures in shape memory alloys [36]. The elastic anisotropy factor is due to Zener [37], which is defined by $A = 2C_{44}/(C_{11}- C_{12})$. According to our calculations the value of $A$ increases from 0.8 to 1.07 as Z increases (Fig. 2), indicating that Sc$_3$InB has a larger anisotropy compared to Sc$_3$InC and Sc$_3$InN.

The ductile-brittle nature of materials is often discussed in terms of elastic constants of the relevant material. The Cauchy's pressure, defined as the difference between the two particular elastic constants ($C_{12} - C_{44}$) is considered to serve as an indication of ductility: if the pressure is positive (negative), the material is expected to be ductile (brittle) [38]. Here the Cauchy's pressures of Sc$_3$In$X$ phases are all negative, which is a clear indication for the compounds to be brittle. Another index of ductility is the Pugh's ratio $G/B$ ratio [39]. The high (low) $G/B$ ratio is associated with the brittle (ductile) nature of the materials. The critical number which separates the ductile and brittle is ~ 0.5. The $G/B$ ratio for Sc$_3$InB, Sc$_3$InC and Sc$_3$InN are 0.71, 0.83 and 0.78, respectively, which clearly highlights the brittle nature of all these perovskites.

The Debye temperature $\Theta_D$ determines the thermal characteristics of the material. As a matter of fact, a higher $\Theta_D$, would imply a higher thermal conductivity associated with the material. Using the relationship with the mean sound velocity ($v_m$), the Debye temperature ($\Theta_D$) [40] can be evaluated. The calculated $\Theta_D$ for Sc$_3$InB, Sc$_3$InC and Sc$_3$InN are found to be 437, 526 and 512 K, respectively. $\Theta_D$ values for Sc$_3$In$X$ compounds are of similar order when compared with that of Ti$_3$AlN [41]. The relatively high $\Theta_D$ values would indicate the lattices to be rather stiff and of good thermal conductivity. The dependence of the elastic constants is a very important characterization of the crystals with varying temperature and pressure, which will be dealt with in a later section.

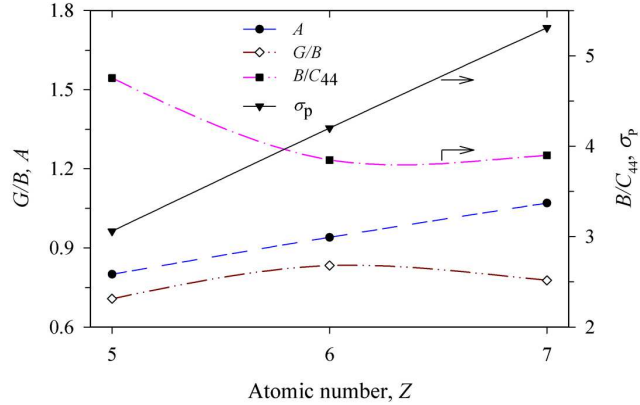

**Fig. 2.** The ratio $G/B$, shear anisotropy $A$, $B/C_{44}$, and $\sigma_P$ of perovskites as a function of atomic number of X atoms.

The movement of a dislocation in a glide plane of the perovskites can be predicted using Peierls stress ($\sigma_P$) [42]. The initial stress value required to initiate such movement is expressed through shear modulus $G$ and Poisson ratio $\nu$ as follows:

$$\sigma_P = \frac{2G}{1-\nu} \exp\left(-\frac{2\pi d}{b(1-\nu)}\right) \qquad (1)$$

where $b$ is the Burgers vector and $d$ is the interlayer distance between the glide planes. For perovskites the Burgers vector $b$ is for <001> plane, and interlayer distance $d$ for 1/2(001) plane [43] (see Table 1). The estimated Peierls stress data for perovskites under consideration are shown in Fig. 2. As can be seen the value progressively increases as we go from B→C→N. We may compare the obtained $\sigma_P$ values with those of a selection of MAX phases Ti$_2$AlC, Cr$_2$AlC, Ta$_2$AlC, V$_2$AlC, and Nb$_2$AlC for which $\sigma_P$ are 0.98, 0.86, 0.78, 0.76, and 0.71 (GPa), respectively) [44]. The reported $\sigma_P$ for rocksalt binary carbide TiC is 19.49 [44], showing the sequence, $\sigma_P$(MAX) < $\sigma_P$(perovskites) < $\sigma_P$(binary carbide). Thus it is clear that in the MAX phases dislocations can move, while this is not the case for the binary carbide. Since the perovskites studied here exhibit an intermediate values of $\sigma_P$, approximately 3-5 times larger than in MAX phases, dislocation movement may still occur here, but not so easily as in MAX phases.

*3.2. Mulliken bonding population and theoretical hardness*

In order to analyze the bonding nature and to obtain theoretical hardness ($H_V$) of Sc$_3$In$X$ ($X$ = B, C and N) we calculated the Mulliken bond populations. Further the relevant formula for the estimation of hardness is given as [45, 46]:

$$H_V = \left[ \prod^\mu \left\{ 740 \left( P^\mu - P^{\mu'} \right) \left( v_b^\mu \right)^{-5/3} \right\}^{n^\mu} \right]^{1/\Sigma n^\mu} \qquad (2)$$

where $P^\mu$ is the Mulliken population of the $\mu$-type bond, $P^{\mu'} = n_{free}/V$ is the metallic population (with $n_{free}$ = number of free electrons), $n^\mu$ is the number of $\mu$-type bond, and $v_b^\mu$ is the bond volume of $\mu$-type bond.

Table 2 shows the calculated bond length, overlap population and the theoretical Vickers hardness. The Mulliken bond populations explain the overlap degree of the electron clouds of two bonding atoms in the crystal. There are in fact both positive and negative values of the population. A positive value indicates the bonding state whereas the negative value refers to the antibonding state. There is no significant interaction between the electronic populations of the two bonding atoms if overlap population is close to zero. The highest value indicates the strong covalency of the chemical bonding. In our calculations, the value of $P^\mu$ of the Sc-$X$ for the three antiperovskite compounds are found in the range from 0.84-0.64, which are larger than the values of the other bond (i.e., Sc-Sc: 0.37−0.15), indicating a stronger covalent nature between Sc and $X$ atoms compared with Sc-Sc bond. We notice that the calculated Sc-B bond in Sc$_3$InB compound is slightly stronger than Sc-C and Sc-N of two antiperovskite compounds Sc$_3$InC and Sc$_3$InN, respectively. The calculated values of theoretical Vickers hardness of Sc$_3$In$X$ are shown in the table.

**Table 2.** Calculated Mulliken bond overlap population of $\mu$-type bond $P^\mu$, bond length $d^\mu$, metallic population $P^{\mu'}$, bond volume $v_b^\mu$ (Å$^3$) and theoretical hardness of $\mu$-type bond $H_v^\mu$ and $H_v$ of Sc$_3$InB, Sc$_3$InC and Sc$_3$InN.

| Compound | Bond | $d^\mu$ (Å) | $P^\mu$ | $P^{\mu'}$ | $v_b^\mu$ (Å$^3$) | $H_v^\mu$ (GPa) | $H_v$ (GPa) |
|---|---|---|---|---|---|---|---|
| Sc$_3$InB | Sc-B | 2.3437 | 0.84 | 0.0677 | 8.97 | 14.75 | 3.88 |
|  | Sc-Sc | 3.3145 | 0.37 | 0.0677 | 25.36 | 1.02 |  |
| Sc$_3$InC | Sc-C | 2.2765 | 0.79 | 0.0370 | 8.22 | 16.64 | 3.81 |
|  | Sc-Sc | 3.2195 | 0.26 | 0.0370 | 23.24 | 0.87 |  |
| Sc$_3$InN | Sc-N | 2.2354 | 0.64 | 0.0097 | 7.779 | 15.25 | 3.03 |
|  | Sc-Sc | 3.1613 | 0.15 | 0.0097 | 21.964 | 0.60 |  |

The bond nature of the compounds can be understood further from the distribution of the total charge density plots in (110) plane (Fig. 3). There are more charges between Sc and *X* atoms, while there are much less charges in the area between the other two atoms (i.e., In-*X*). The charge density distribution of three compounds in (110) plane are almost same but there is a slight difference in the middle area between Sc and *X* atoms. The spherical-like charge density distribution around In atoms and very small charge density in the region between the In atoms and Sc atoms indicate that In-Sc is prone to form the ionic bond nature.

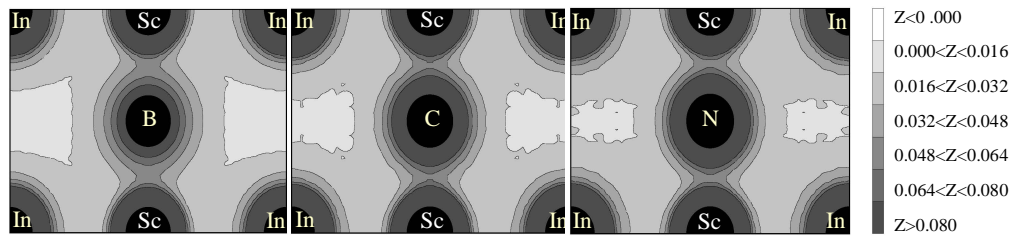

**Fig. 3.** The electron density maps of $Sc_3InX$ in the (110) plane from DFT calculations.

*3.3. Some features of Fermi surface*

Figs. 4 shows the calculated Fermi surfaces of $Sc_3InX$ ($X$ = B, C, N) for the bands which cross the Fermi level. It is to be noted that the important contribution of the density of states near the Fermi level (not shown here) in all the three perovskites is from Sc *d*-states, with a small admixture of B *p*, C *p* or N *p*-states. As Sc 3*d* electrons are mainly involved around the Fermi level, the metallicity in the three perovskites is determined by these electrons.

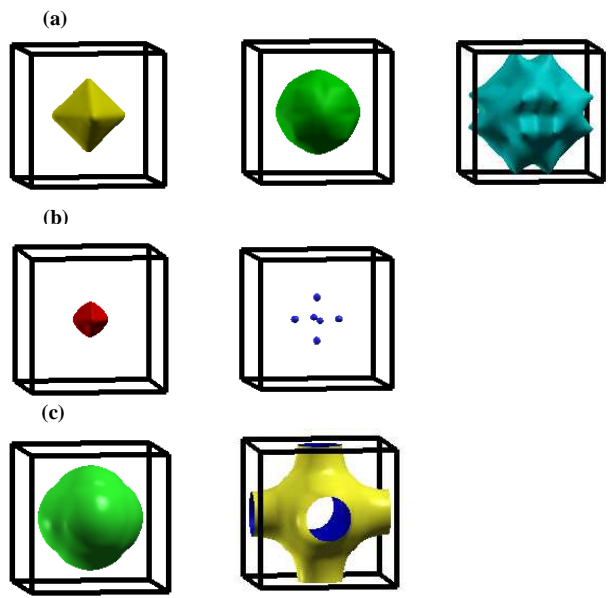

**Fig. 4.** Fermi surfaces of (a) $Sc_3InB$, (b) $Sc_3InC$, and (c) $Sc_3InN$.

We observe from these figures that the topology of Fermi surface changes by replacing B with C or N. The Fermi surface of Sc$_3$InB contains both hole- and electron-like topology, which are centered at the zone center Γ point as well as along Γ→X direction ('X' point is the center of the face of the Brillouin zone). The two large electron sheets for Sc$_3$InB show close similarity to a free-electron Fermi spheres with closely delineated surfaces. The band structure calculations (not shown here) reveal that when we replace B by C or N, the bands crossing the Fermi level at Γ and Γ→X direction are shifted downwards. This leads to an electron pocket at the Γ point in Sc$_3$InN, in contrast to a hole pocket in Sc$_3$InC. In fact the Fermi surface of Sc$_3$InC contains a smaller hole pocket at Γ point, when compared to Sc$_3$InB. Likewise, we also notice a shift of bands along Γ→X direction leading to an interconnected tubular like electron sheet in the case of Sc$_3$InN, which is less pronounced in the case of Sc$_3$InC.

*3.4. Thermodynamic properties at elevated temperature and pressure*

The thermodynamic properties of Sc$_3$In$X$ are investigated by using the quasi-harmonic Debye model, as implemented in the Gibbs program, the detailed description of which can be found in literature [33]. Using the procedure the bulk modulus, specific heats, Debye temperature and volume thermal expansion coefficient at different temperatures and pressures are then calculated for the first time. To start with we utilized *E-V* data obtained from the third-order Birch-Murnaghan equation of state [34] using zero temperature and zero pressure equilibrium values, $E_0$, $V_0$, $B_0$, based on DFT method within the generalized gradient approximation. The thermodynamic properties at finite-temperature and finite-pressure can then be obtained using the model. The non-equilibrium Gibbs function G$^*$(*V*; *P*, *T*) can be written in the form [33]:

$$G^*(V;P,T) = E(V) + PV + A_{vib}[\Theta(V);T] \qquad (2)$$

where *E(V)* is the total energy per unit cell, *PV* corresponds to the constant hydrostatic pressure condition, *Θ(V)* is the Debye temperature, and $A_{vib}$ is the vibrational term. The last term can be expressed using the Debye model of the phonon density of states as [33]:

$$A_{vib}(\Theta,T) = nkT\left[\frac{9\Theta}{8T} + 3\ln(1-\exp(-\Theta/T)) - D\left(\frac{\Theta}{T}\right)\right] \qquad (3)$$

where *D(Θ/T)* represents the Debye integral, and *n* is the number of atoms per formula unit.

The Gibbs function can now be minimized with respect to volume *V* to obtain the thermal equation of state *V(P, T)* and the chemical potential *G(P, T)* of the corresponding phase. Other macroscopic properties like the bulk modulus, Debye temperature specific heats, and volume thermal expansion coefficient can now be evaluated as a function of *P* and *T* from standard thermodynamic relations [33].

Fig. 5(a) shows the temperature dependence of isothermal bulk modulus of Sc$_3$In$X$ phases at *P* = 0 GPa. The zero pressure theoretical results for Sc$_3$InC due to Haddadi *et al.* [20] shown as (+) in the figure are seen to decrease at a faster rate. The bulk modulus increases as *Z* increases, and as the temperature increases the bulk modulus decreases slowly and the rate of decrease is nearly same for all the compounds under consideration. Further the bulk modulus, signifying the average strength of the coupling between the neighboring atoms, increases with pressure at a given temperature (see inset of Fig. 5 (a)) and decreases with temperature at a given pressure, which is consistent with the trend of volume of the perovskites.

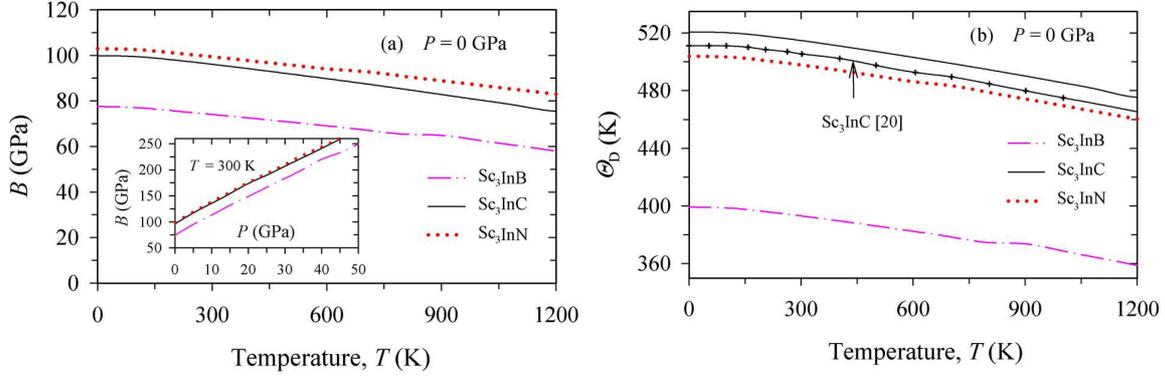

**Fig. 5.** The temperature dependence of (a) bulk modulus and (b) Debye temperature of $Sc_3InX$ at $P = 0$ GPa in comparison with results (+) for $Sc_3InC$ [20]. The inset to the left figure shows the pressure dependence of $B$ at 300 K.

Fig. 5 (b) displays temperature dependence of Debye temperature $\Theta_D$ for the three perovskites at zero pressure along with the theoretical result due to Haddadi *et al.* [20] for only $Sc_3InC$ compound. Our $\Theta_D$ value at $P = 0$ GPa, $T = 0$ K for $Sc_3InC$ is 521 K and that of Haddadi *et al.* [20] is 511 K. One observes that $\Theta_D$, smallest for $Sc_3InB$ phase, decrease slightly in a non-linear way with temperature for all the perovskites. Further pressure variation of $\Theta_D$ (not shown) also shows a non-linear increase. The variation of $\Theta_D$ with pressure and temperature reveals that the thermal vibration frequency of atoms in the perovskites changes with pressure and temperature.

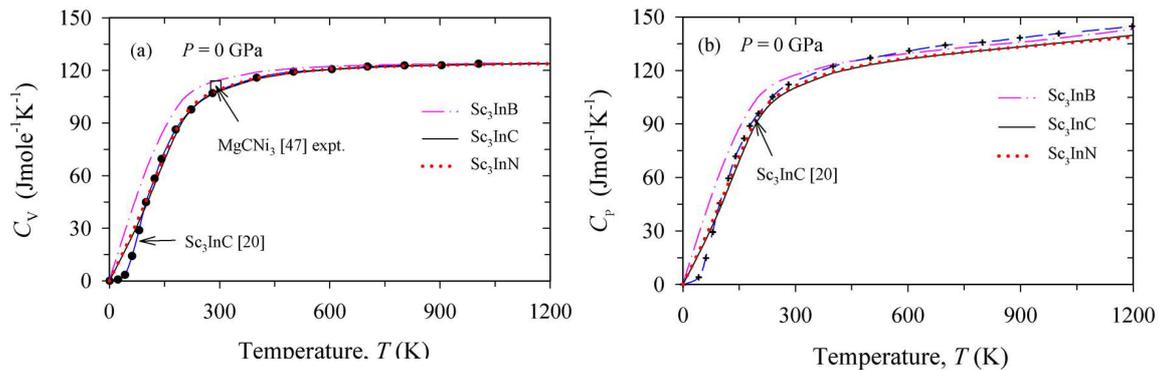

**Fig. 6.** The temperature dependence of (a) constant-volume and (b) constant-pressure specific heat capacities $C_V$, $C_P$ of $Sc_3InX$ ($X = B$, C and N) in comparison with available theoretical result [20].

Fig. 6 shows the temperature dependence of constant-volume and constant-pressure specific heat capacities $C_V$, $C_P$ of $Sc_3InX$ ($X = B$, C and N). We compare our results with the available theoretical results for $Sc_3InC$ [20]. There is no experimental value of specific heat capacities. But we show a lone experimental data point for the superconducting $MgCNi_3$ at room temperature [47]. The heat capacities increase with increasing temperature, because phonon thermal softening occurs when the temperature increases. The difference between $C_P$ and $C_V$ for Sc$_3$In$X$ is given by $C_P - C_V = \alpha_V^2(T) BTV$, which is due to the thermal expansion caused by anharmonicity effects. In the low temperature limit, the specific heat exhibits the Debye $T^3$ power-law behavior and at high temperature ($T > 400$ K) the anharmonic effect on heat capacity is suppressed, and $C_V$ approaches the classical asymptotic limit of $C_V = 3nNk_B = 124.7$ J/mol.K for Sc$_3$In$X$ ($X$ = B, C and N). These results show the fact that the interactions between ions in Sc$_3$In$X$ have great effect on heat capacities especially at low temperatures.

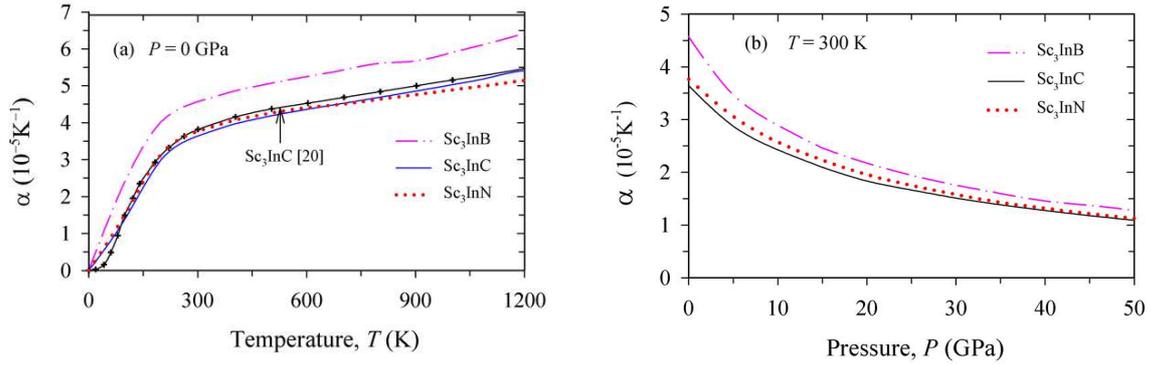

**Fig. 7.** The volume thermal expansion coefficient $\alpha_V$ as a function of (a) temperature and (b) pressure for Sc3In$X$ ($X$ = B, C and N) compared with available results.

Figs. 7 (a) and (b) show the volume thermal expansion coefficient $\alpha_V$ of Sc$_3$In$X$ perovskites as a function of temperature and pressure, respectively. In the absence of experimental data, the available theoretical results of Sc$_3$InC due to Haddadi *et al.* [20] are shown for comparison. The expansion coefficient increases rapidly especially at temperature below 300 K, whereas it gradually tends to a slow increase at higher temperatures. On the other hand at a constant temperature, the expansion coefficient decreases strongly with pressure. It is well-known that the thermal expansion coefficient is inversely related to the bulk modulus of a material.

*3.5. Optical properties*

The compound with cubic symmetry requires only one dielectric tensor component to characterize the linear optical properties. The imaginary part of the dielectric function $\varepsilon_2(\omega)$ is calculated using the expression given in Ref. [47]. The real part $\varepsilon_1(\omega)$ is derived from the imaginary part by the Kramers-Kronig transform. From the real and imaginary parts of the dielectric function one can calculate other spectra, such as refractive index, absorption spectrum, loss-function, reflectivity and conductivity (real part) using the expressions given in Ref. [47].

The optical functions of Sc$_3$In$X$ as a function of photon energies up to 20 eV for polarization vectors [100] are displayed in Fig. 8. In all cases reported here we use a 0.5 eV Gaussain smearing, which smears out the Fermi level in such a way that $k$-points will be more effective on the Fermi surface. The calculations only include interband exciatations. It is known that in metal and metal-like systems there are intraband contributions from the conduction electrons mainly in the low-energy infrared part of the spectra. Electronic band structures (not shown here) indicate that the inverse-perovskites considered here are metallic in nature and further their bonding is characterized by a mixture of ionic-covalent type [20-21]. It is thus necessary to include an empirical Drude term to the dielectric function [48, 49]. A Drude term with plasma frequency 3 eV and damping (relaxation energy) 0.05 eV was used.

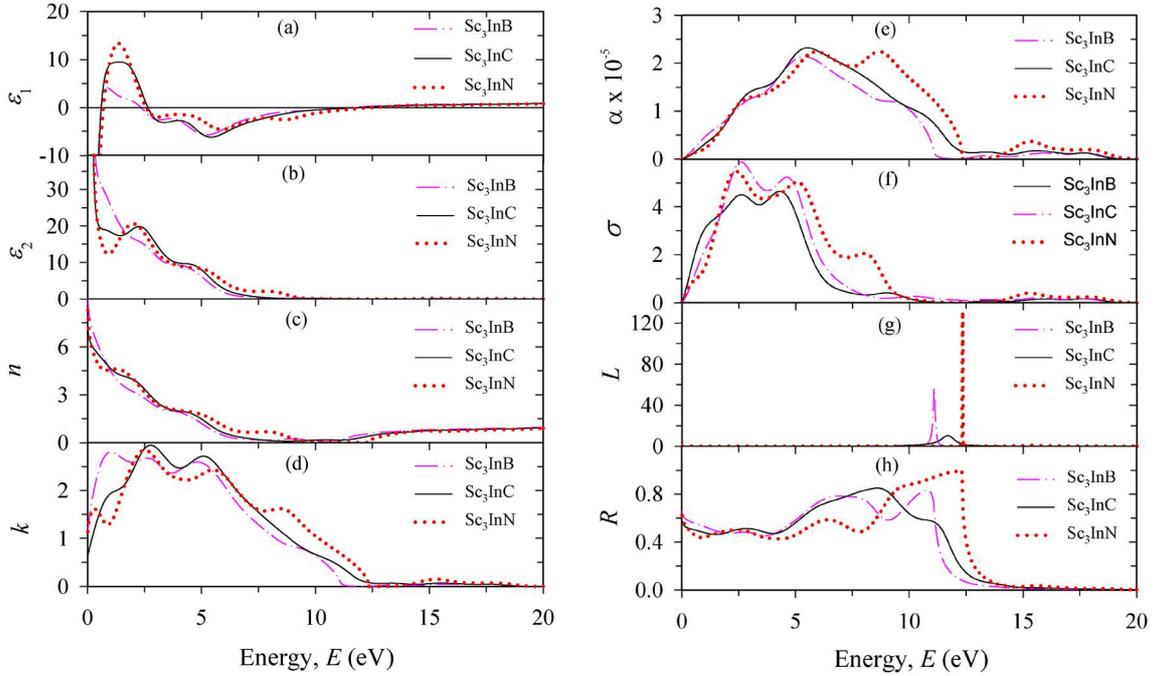

**Fig. 8.** Energy dependent (a) real part of dielectric function, (b) imaginary part of dielectric function, (c) refractive index, (d) extinction coefficient, (e) absorption, (f) real part of conductivity, (g) loss function, and (h) reflectivity of Sc$_3$In$X$ along [100] direction.

The imaginary and real parts of the dielectric function are displayed in Fig. 8 (a, b). It is observed that the real part $\varepsilon_1$ goes through zero from below at about 11.3, 11.8, and 12.5 eV and the imaginary part $\varepsilon_2$ approaches zero from above at about 9.8, 8.8, and 11.1 eV for Sc$_3$InB, Sc$_3$InC and Sc$_3$InN, respectively. Metallic reflectance characteristics are exhibited in the range of $\varepsilon_1 < 0$. The peak of the imaginary part of the dielectric function is related to the electron excitation. For the imaginary part, $\varepsilon_2$, the peak for < 0.8 eV is due to the intraband transitions. It is clear from the figure that $\varepsilon_2(\omega)$ shows single peak at ~ 2 eV (Sc$_3$InN) and at ~ 2.3 eV (Sc$_3$InC). The refractive index and extinction coefficient are illustrated in Fig. 8 (c) and (d), respectively. The calculated static refractive index $n(0)$ is found to be 9.4, 7.2, and 8.3, for Sc$_3$InB, Sc$_3$InC, and Sc$_3$InN, respectively.

The absorption coefficient spectra of the perovskites are displayed in Fig. 8 (e). As is evident from band structures (not shown here) the materials have no band gap and hence the photoconductivity starts with zero photon energy for each of the phases as shown in Fig. 8 (f). These spectra have several maxima and minima within the energy range studied. The photoconductivities and hence electrical conductivities of materials increase as a result of absorbing photons. The maximum optical conductivity [in (mΩcm)$^{-1}$] of 4.64 (Sc$_3$InB), 5.47 (Sc$_3$InC), and 5.85 (Sc$_3$InN) was observed at photon energy of ~ 4.25, 2.36, and 2.53 eV, respectively.

The electron energy loss spectra $L(\omega)$ and reflectivity spectra $R(\omega)$ of the three inverse-perovskites as a function of photon energy are shown in Fig. 8 (g) and Fig. 8 (h), respectively. The function, $L(\omega)$, unfolds the energy loss of a fast electron passing through a material. It is observed that $L(\omega)$ shows three sharp peaks one for each phase at 11, 11.7, and 12.3 eV for Sc$_3$InB, Sc$_3$InC, and Sc$_3$InN, respectively. This peak represents the feature that is associated with plasma resonance, and the corresponding frequency is called bulk plasma frequency. Further these peaks correspond to irregular edges in the reflectivity spectrum (Fig. 8 (h)), and hence an abrupt reduction occurs at these peaks values in the reflectivity spectrum and it correlates with the zero crossing of $\varepsilon_1(0)$ with small $\varepsilon_2(0)$, shown in Fig. 8 (a, b). The reflectivity values of the three perovskites are high between 0 and ~12.3 eV photon energy, reaching maximum of ~99%. The reflectivity is thus much higher than those in the superconducting antiperovskite MgCNi$_3$ [47]. It implies that Sc$_3$In$X$ can all be used as good coating materials in the ultraviolet region.

## 4. Conclusion

First-principles calculations based on DFT have been used to study the mechanical, thermodynamic and optical properties of Sc$_3$InB, Sc$_3$InC and Sc$_3$InN. The materials are elastically anisotropic and brittle. The estimated values of Peierls stress indicate that dislocation movement may follow in all the perovskites considered but with much reduced occurrences compared to MAX phases.

The Mulliken bonding population and charge density maps show stronger covalent nature between Sc and $X$ atoms compared with Sc-Sc bond. We notice that the calculated Sc-B bond in Sc$_3$InB compound is slightly stronger than Sc-C and Sc-N of two antiperovskite compounds Sc$_3$InC and Sc$_3$InN, respectively. The Vickers hardness values of Sc$_3$In$X$ are also predicted to be in the range 3.03 – 3.88 GPa. The Fermi surfaces of Sc$_3$In$X$ contain both hole- and electron-like topology and these are seen to change by replacing B with C or N.

The finite-temperature ($\leq$ 1200 K) and finite-pressure ($\leq$ 50 GPa) thermodynamic properties, *e.g.* bulk modulus, specific heats, thermal expansion coefficient, and Debye temperature are all obtained through the quasi-harmonic Debye model, which considers the vibrational contribution, and the results are analysed. The variation of $\Theta_D$ with temperature and pressure reveals the changeable vibration frequency of the particles in Sc$_3$In$X$ phases. The heat capacities increase with increasing temperature, which shows that phonon thermal softening occurs when the temperature increases. From the analysis of optical functions for the polarization vectors [100], it is found that the reflectivity is high in IR-UV regions and it implies that Sc$_3$In$X$ ($X$ = B, C and N) materials can be used as potential coating materials.